\documentclass[epj]{svjour}
\usepackage{latexsym,graphicx,url}
\usepackage[american]{babel}

\newcommand{\q}[2]{\ensuremath{#1\ \mathrm{#2}}}

\newcommand{\degC}{\ensuremath{\left.^\circ\mathrm{C}\right.}}
\newcommand{\is}[2]{\ensuremath{\left.^{#1}\mathrm{#2}\right.}}
\newcommand{\nzero}{\ensuremath{n_\mathrm{eq}}}
\newcommand{\Nzero}{\ensuremath{N_\mathrm{eq}}}
\newcommand{\Ntrap}{\ensuremath{N_\mathrm{t}}}
\newcommand{\Nvap}{\ensuremath{N_\mathrm{v}}}
\newcommand{\phiesc}{\ensuremath{\phi_\mathrm{esc}}}
\newcommand{\phichem}{\ensuremath{\phi_\mathrm{chem}}}
\newcommand{\phidec}{\ensuremath{\phi_\mathrm{dec}}}
\newcommand{\vavg}{\ensuremath{\bar{v}}}
\newcommand{\vcapture}{\ensuremath{v_\mathrm{c}}}
\newcommand{\tauesc}{\ensuremath{\tau_\mathrm{esc}}}
\newcommand{\tausto}{\ensuremath{\tau_\mathrm{sto}}}
\newcommand{\taudec}{\ensuremath{\tau_\mathrm{dec}}}
\newcommand{\tauloss}{\ensuremath{\tau_\mathrm{loss}}}
\newcommand{\Rlaser}{\ensuremath{R_b}}

\begin{document}
\title{Experimental study of vapor-cell magneto-optical traps for
  efficient trapping of radioactive atoms}

\author{S.~N.~Atutov\inst{1,2}%
\thanks{Corresponding author. \email{atutov@fe.infn.it}.
  \emph{Permanent address:} Institute for Automation and Electrometry
  Sib. RAS, Koptuga 1, 630090 Novosibirsk, Russia} \and
        R.~Calabrese\inst{1,2} \and
        A.~Facchini\inst{1} \and
        G.~Stancari\inst{2} \and
        L.~Tomassetti\inst{1,2}}
\authorrunning{S.~N.~Atutov et al.}

\institute{ Dipartimento di Fisica, Universit\`a di Ferrara, Via
  Saragat 1, I-44100 Ferrara FE, Italy \and
  Istituto Nazionale di
  Fisica Nucleare, Sezione di Ferrara, Via Saragat 1, I-44100 Ferrara FE, Italy}

\date{Received: date / Revised version: date}

\abstract{We have studied magneto-optical traps (MOTs) for efficient
  on-line trapping of radioactive atoms. After discussing a model of
  the trapping process in a vapor cell and its efficiency, we present
  the results of detailed experimental studies on Rb MOTs.  Three
  spherical cells of different sizes were used. These cells can be
  easily replaced, while keeping the rest of the apparatus unchanged:
  atomic sources, vacuum conditions, magnetic field gradients, sizes
  and power of the laser beams, detection system.  By direct
  comparison, we find that the trapping efficiency only weakly depends
  on the MOT cell size. It is also found that the trapping efficiency
  of the MOT with the smallest cell, whose diameter is equal to the
  diameter of the trapping beams, is about 40\% smaller than the
  efficiency of larger cells.  Furthermore, we also demonstrate the
  importance of two factors: a long coated tube at the entrance of the
  MOT cell, used instead of a diaphragm; and the passivation with an
  alkali vapor of the coating on the cell walls, in order to minimize
  the losses of trappable atoms.  These results guided us in the
  construction of an efficient large-diameter cell, which has been
  successfully employed for on-line trapping of Fr isotopes at INFN's
  national laboratories in Le\-gna\-ro, Italy.%
   }

\maketitle

\section{Introduction}
\label{sec:intro}

There are several fields of research for which trapped radioactive
atoms can be a useful tool.  Precise atomic spectroscopy is a test of
relativistic quan\-tum-me\-chan\-i\-cal many-body predictions. Besides
their intrinsic interest, these studies are the basis for testing the
electroweak model via atomic parity non-conservation (APNC). The
electron-nucleon interaction can be probed by measuring the weak
charge of the nucleus at low momentum transfers, which is
complementary to measurements at high energy; and the measurement of
nuclear anapole moments is a unique tool to access weak
nucleon-nucleon interactions. The direct study of time-reversal
symmetry through the search for permanent electrical dipole moments
(EDMs) might also be accessible with trapped atoms.  Recent reviews of
this field can be found in
Refs.~\cite{Ginges:PR:2004,Guena:MPLA:2005}.

Successful trapping of short-lived radioactive atoms was de\-mon\-stra\-ted
at Berkeley~\cite{Lu:PRL:1994}. In this experiment, \is{21}{Na} atoms
were trapped for the purpose of testing the $V-A$ structure of the
electroweak interaction by performing precise measurements of $\beta$
decays.  Almost simultaneously, \is{79}{Rb} atoms were trapped at SUNY
Stony Brook~\cite{Gwinner:PRL:1994}. This group was also the first to
trap francium~\cite{Simsarian:PRL:1996} and to perform extensive
spectroscopy on these atoms~\cite{Gomez:RPP:2006}.  The group at JILA
Boulder trapped \is{221}{Fr} generated by the decay chain of
\is{229}{Th} and performed several spectroscopic
measurements~\cite{Lu:PRL:1997}.

Present activities with radioactive atoms in magneto-optical traps
include parity violation in $\beta$ decays of Rb and Cs isotopes at
LANL~\cite{Crane:PRL:2001} and Na isotopes by the TRI$\mu$P team at
KVI (Netherlands)~\cite{Traykov:NIMB:2008}; the study of anapole
moments in francium at TRIUMF (Canada)~\cite{Gwinner:HI:2006}; and EDM
searches in Ra at ANL (USA)~\cite{Guest:PRL:2007} and
KVI~\cite{De:arXiv:2008}, and in Fr at RCNP-CYRIC in
Japan~\cite{Sakemi:private:2007}.  Our group built the first European
facility for the production and trapping of francium at INFN's
national laboratories in Le\-gna\-ro,
Italy~\cite{Atutov:NPA:2004,Stancari:NIMA:2006,Stancari:NIMA:2008,%
  Stancari:EPJST:2007}.  Our long-term goal is to measure parity
violation in the atomic transitions of francium.

Francium is the heaviest alkali metal. Its large and highly-charged
nucleus enhances APNC and EDM effects. At the same time, its atomic
structure is relatively simple, and precise calculations are possible.
Francium has several isotopes with lifetimes $> \q{1}{min}$.  They are
suitable for trapping and can be compared to estimate nuclear effects.

There are no stable Fr isotopes, but traps can partly compensate for
their scarcity. In particular, the magneto-optical trap can cool the
atom cloud to temperatures in the millikelvin range and this, due to
suppression of the Doppler broadening of atomic lines, greatly
increases the intensity of atomic flu\-o\-res\-cence and
ab\-sorp\-tion of trapped atoms. This makes these traps suitable for
high-resolution, Doppler-free or nonlinear spectroscopy.

The magneto-optical trap also allows one to perform measurements in
the pulsed regime. Atoms can be collected on the cell walls or on the
surface of the neutralizer. They are then periodically released and
accumulated in the trap.  The perturbing effects of electromagnetic
fields are eliminated by periodically turning off the trap's magnetic
field and laser beams. Measurements can be done during the field-free
expansion of the atom cloud.  Moreover, in the pulsed regime, the
lock-in technique for the detection of small signals can be employed.

Obviously, an efficient optical trapping process is of great
importance for the creation of large samples of radioactive
atoms. Improvements in the collection efficiency of a MOT is a key
consideration for experiments featuring very weak atomic fluxes.  The
trapping efficiency depends upon several factors, such as laser power,
laser beam size, quality of coating, pumping port design, magnetic
field gradient, etc.

It is also believed that, in order to obtain maximum efficiency, one
should design a cell with a large ratio between the volume occupied by
the laser beams and the total cell
volume~\cite{Lu:PRL:1997,Stephens:PRL:1994,Aubin:RSI:2003}.  For
instance, for an available trapping laser power of 1~W and for a
saturation power of \q{3\mbox{--}7}{mW/cm^2}, the cell diameter must be
of the order of 5~cm for full overlap of laser beams and cell volume.
Some serious problems arise with this
kind of cell. First of all, a small cell makes it difficult to insert
a hot piece of metal (neutralizer) inside the trapping volume. The
neutralizer is necessary when radioactive isotopes are transported as
ions. A hot neutralizer inside the trap volume can damage the wall
coating and degrade vacuum conditions.  A possible solution is to
place the neutralizer outside the cell's pumping port, but this
usually reduces the collection efficiency dramatically.  Moreover, a
small cell suffers from high levels of stray light, which make it
difficult to directly detect low levels of fluorescence.

A possible dependence of the trapping efficiency on the MOT cell size
for a given laser beam radius is the main object of our experimental
study.  We trapped Rb atoms in a magneto-optical trap with three
different vapor cells.  These cells have different volumes but exactly
the same pumping port, which consists of a coated glass tube.  In all
these experiments, the same source of atoms and optical detection
system were used. We also present the results of a study on the
variation of the density of trappable atoms in the cell as a function
of port tube length; and we demonstrate the importance of passivation
of the coating on the cell walls with an alkali vapor in order to
minimize the loss rate of trappable atoms.  These experimental studies
are preceded by the discussion of a model of the trapping process and
by the definition of the relevant quantities.

\section{Loading of a vapor cell}
\label{sec:laading}

Let us consider a spherical cell with radius~$R$ and with an entrance
or exit port in the shape of a cylindrical tube with internal
radius~$r$ and length~$l$.  The cell is connected to a vacuum pump through
the exit port and a valve. The surfaces of the cell, port, and valve
are covered with a non-stick coating. Radioactive ions from a beam
trasport line are injected into the cell and impinge on the
neutralizer, a hot metal plate placed at the far end of the cell.  The
trap's laser beams and magnetic field are turned off.

After the radioactive ion beam is switched on, the ions come inside
the cell and impinge on the neutralizer, stick to its surface for a
short time, become neutralized and are finally desorbed and released
into the cell volume. Atoms can also be injected directly into the
cell in neutral form.  In both cases, atoms start to fill the cell and
to saturate its walls, and the density of the vapor starts to
increase.  The density of atoms in the cell reaches an equilibrium
value~\nzero\ when the sum of all loss rates ---
the leaking or escaping of atoms from the cell through the port tube \phiesc,
their chemical adsorption on the cell walls \phichem\ and their
radioactive decay \phidec\ --- becomes equal to the flux~$I$ of atoms
into the cell:
\begin{equation}
I = \phiesc + \phichem + \phidec.
\label{eq:equilib}
\end{equation}

In our case, \phiesc\ can be calculated from the conductance~$K$ of the
port tube in the molecular-flow regime multiplied by the the density
difference~$\Delta n \simeq \nzero$:%
\begin{equation}
\phiesc = K \cdot \Delta n = \frac{2 \pi r^3 \vavg \nzero}{3 l},
\label{eq:phiesc}
\end{equation}
where $\vavg = \sqrt{8kT/(\pi m)}$ is the average atomic thermal
velocity at temperature~$T$, and $m$ is the mass of the atom.

The loss rate \phichem\ is the flux of atoms towards the internal
surface of the cell walls. The atoms are absorbed with probability $1
/\chi$ by chemiosorption on the coating. The parameter~$\chi$ is also
interpreted as the average number of bounces it takes to adsorb atoms
on the surface of the coating.  The loss rate by chemisorption can be
expressed in the following form:
\begin{equation}
\phichem = \frac{4 \pi R^2 \vavg \nzero}{\chi}.
\label{eq:phichem}
\end{equation}
Here we neglect the chemical loss of atoms in the neutralizer.

The loss rate due to radioactive decay is:
\begin{equation}
\phidec = \frac{\Nzero}{\taudec} = \frac{\nzero V}{\taudec},
\end{equation}
where~$\Nzero = \nzero V$ is the total number of atoms in the cell, $V$ is the
cell volume $4 \pi R^3 / 3$, and $\taudec$ is the radioactive
lifetime. In a coated cell with a sufficiently hot neutralizer, the
sticking time of atoms to the coating and to the surface of the
neutralizer is much shorter than their radioactive lifetime. The
radioactive loss of atoms on the cell walls or on the neutralizer can
therefore be neglected.

We note that in this model the requirement for the sticking time of
atoms to the coating is much less strict than the one discussed in
Ref.~\cite{Stephens:JAP:1994}: for obtaining a high trapping
efficiency, it is sufficient that the sticking time be small compared
with the radioactive lifetime.

The equilibrium condition (Eq.~\ref{eq:equilib}) becomes
\begin{equation}
I = \frac{2 \pi r^3 \vavg \nzero}{3 l} +
    \frac{4 \pi R^2 \vavg \nzero}{\chi} +
    \frac{V \nzero}{\taudec},
\end{equation}
and one can write the total number of atoms in the cell at
equilibrium as follows:
\begin{eqnarray}
\Nzero & = & I \cdot \left[ \frac{\vavg}{2l}\left(\frac{r}{R}\right)^3 +
                \frac{3 \vavg}{\chi R} +
                \frac{1}{\taudec} \right]^{-1} \label{eq:Neq} \\
       & \equiv & I \cdot \left[ \frac{1}{\tauesc} +
                        \frac{1}{\tausto} +
                        \frac{1}{\taudec} \right]^{-1}. \nonumber
\end{eqnarray}
The first term between square brackets in Eq.~\ref{eq:Neq} is the
inverse escape time~\tauesc, which represents the average time it
takes to lose atoms through the pumping port:
\begin{equation}
\tauesc \equiv \frac{2 l}{\vavg} \left( \frac{R}{r} \right)^3.
\label{eq:tauesc}
\end{equation}
The second term is the inverse
storage time~\tausto\ of atoms inside a closed cell before being lost
to chemisorption on the cell walls:
\begin{equation}
\tausto \equiv \frac{\chi R}{3 \vavg}.
\label{eq:tausto}
\end{equation}
In the steady-state regime, the total number of atoms in the cell can
be written in the following compact form:
\begin{equation}
\Nzero = \frac{I}{W},
\label{eq:IW}
\end{equation}
where~$W$ is the total loss rate of trappable atoms, defined as follows:
\begin{equation}
W \equiv \frac{1}{\tauloss} \equiv
  \frac{1}{\tauesc} + \frac{1}{\tausto} + \frac{1}{\taudec},
\label{eq:W}
\end{equation}
and \tauloss\ is the total loss time.

In a typical trapping experiment with ra\-dio\-ac\-tive atoms in a coated
cell, the escape time is much shorter than both the storage time and
the radioactive lifetime. Under these conditions, the contribution of
the storage and decay times can be neglected, and the total number of
atoms at equilibrium becomes
\begin{equation}
\Nzero = I \cdot \tauesc =
  I \cdot \frac{2 l}{\vavg} \left(\frac{R}{r}\right)^3,
\label{eq:Itauesc}
\end{equation}
and the density of trappable atoms in the cell is
\begin{equation}
\nzero \equiv \frac{\Nzero}{V} = \frac{3 l I}{2 \pi \vavg r^3}.
\label{eq:neq}
\end{equation}
For instance, in the case of francium, we have $\vavg = \q{1.7\times
  10^4}{cm/s}$ and $I = \q{10^6}{atoms/s}$. For $l = \q{16}{cm}$ and
$r = \q{1}{cm}$, we expect the steady-state density to be $\nzero =
\q{400}{atoms/cm^3}$.  It is interesting to note that the steady-state
density is proportional to the incoming flux~$I$, to the length~$l$ of
the cell port tube, and to the inverse cube of the port tube
radius~$r$; and it is independent of the cell radius~$R$.

Let us consider the case of a coated cell whose inlet is a diaphragm
($l \ll r$) instead of a port tube.  Also in this case, the escape
time (Eq.~\ref{eq:tauesc}) is small compared to the storage time
(Eq.~\ref{eq:tausto}), and the steady-state density does not depend on
cell size.  A high density of trappable atoms is an important starting
point for obtaining a large trap population.  In the case of a
diaphragm, one expects the density to be much smaller with respect to
the case of a port tube of the same diameter. In the latter, ions are
injected into the cell by a fast and free ballistic flight, and atoms
leak out by diffusion, which is relatively slow.

In the case of an extremely small entrance port~\cite{Lu:PRL:1997} or
of an uncoated cell, the storage time is smaller than both the escape
time and the radioactive lifetime. The loss of atoms is dominated by
chemisorption on the cell walls, and the equilibrium density can be
expressed as follows:
\begin{equation}
\nzero = \frac{\chi I}{4 \pi R^2 \vavg}.
\end{equation}
In this case, the steady-state density is proportional to the product
of the incoming flux and number of atom bounces inside the cell, and
it is inversely proportional to the internal area of the cell surface.
In this particular case, the size of the cell must be kept as small as
possible in order to optimize the steady-state density of trappable
atoms, the number of trapped atoms and the trapping efficiency
(discussed below).

\section{Trapping process in a coated cell}
\label{sec:trapping}

Let us consider the trapping process in a magneto-optical trap with a
coated cell containing a vapor of trappable atoms. The trapping
process can be modeled according to Ref.~\cite{Lu:PRL:1997}. The time
evolution of the number of trapped atoms~\Ntrap\ and the number of
trappable atoms in the vapor~\Nvap\ depends on three parameters: $L$,
the loading rate of atoms from the vapor to the trap; $C$, the
collisional loss rate of atoms from the trap to the vapor due to
collisions with the rest gas; and $W$ (mentioned above), the total
loss rate of atoms from the vapor. In differential form, this model
can be expressed as follows:
\begin{equation}
\left\{
\begin{array}{l}
\dot{\Ntrap} = L \Nvap - C \Ntrap \\
\dot{\Nvap} = C \Ntrap - L \Nvap - W \Nvap + I
\end{array}
\right.
\label{eq:trapmodel}
\end{equation}
Here we neglect the loss of trapped atoms due to their collisions with
other atoms in the trap.

At equilibrium, the number of trapped atoms is
\begin{equation}
\Ntrap = \frac{L I}{C W},
\label{eq:Ntrapeq}
\end{equation}
and the total number of trappable atoms in the vapor is
\begin{equation}
\Nvap = \Nzero = \frac{I}{W}.
\label{eq:Nvapeq}
\end{equation}
Taking Eqs.~\ref{eq:Ntrapeq} and~\ref{eq:Nvapeq} into account, one obtains:
\begin{equation}
\Ntrap = \frac{L \Nzero}{C} = \frac{L V \nzero}{C}.
\label{eq:Ntrapeq2}
\end{equation}
According to Ref.~\cite{Stephens:PRL:1994}, the combination $L V
\nzero$ in Eq.~\ref{eq:Ntrapeq2} is the trap capture rate $\gamma
\Rlaser^2 \nzero$, where \Rlaser\ is the radius of the trapping laser
beams and~$\gamma$ a parameter that depends upon the thermal velocity
of trappable atoms \vavg\ and the trap capture velocity \vcapture. The
number of trapped atoms can be written in the following form:
\begin{equation}
\Ntrap = \frac{\gamma \Rlaser^2 \nzero}{C}.
\label{eq:Ntraplaser1}
\end{equation}
Taking Eq.~\ref{eq:neq} into account, we finally obtain:
\begin{equation}
\Ntrap = \frac{\gamma \Rlaser^2}{C} \frac{3 l I}{2 \pi \vavg r^3}.
\label{eq:Ntraplaser2}
\end{equation}
One can see from Eqs.~\ref{eq:Ntraplaser1} and~\ref{eq:Ntraplaser2}
that the steady-state number of trapped atoms is proportional to the
density of the trappable atoms in the vapor; and that it is
proportional to the flux, to the radius of the laser beams squared, to
the length of the pumping tube, and inversely proportional to the
collision rate of cold atoms with the rest gas and to the radius of
the pumping tube cubed.

One can define a (dimensional) trapping efficiency~$\eta$ as the ratio
between the number of trapped atoms at equilibrium and the incoming
atomic flux:
\begin{equation}
\eta \equiv \frac{\Ntrap}{I} =
  \frac{\gamma \Rlaser^2}{C} \frac{3 l}{2 \pi \vavg r^3}.
\label{eq:eff}
\end{equation}
According to this model, the
trapping efficiency does not depend on the cell size,
unless this dependence is hidden in the constant~$\gamma$.

\section{Experimental setup}
\label{sec:setup}

\begin{figure}
\begin{center}
\resizebox{\columnwidth}{!}{\includegraphics{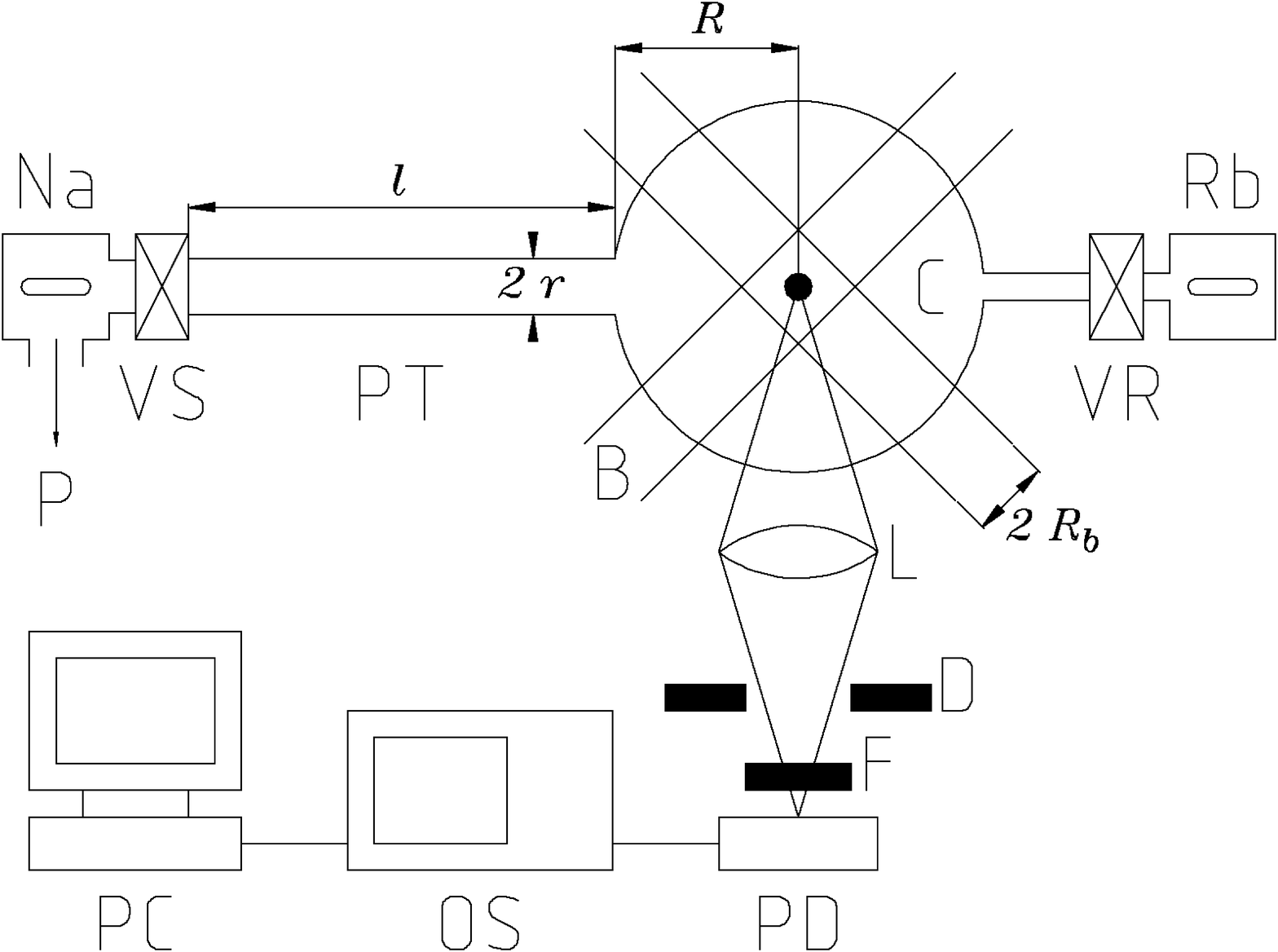}}
\end{center}
\caption{Schematic diagram of the apparatus: atomic sources (Na and
  Rb); valves (VS and VR); port tube (PT), length~$l$ and inner
  radius~$r$; trapping lasers beams (B), radius~\Rlaser; trap cell
  (C), radius~$R$; collection lenses (L); iris diaphragm (D);
  interferential filter (F); photodetector (PD); digital oscilloscope
  with lock-in amplifier (OS); computer (PC); pumping port (P).}
\label{fig:apparatus}
\end{figure}

Our MOT consists of a spherical glass cell with a glass port tube
(Fig.~\ref{fig:apparatus}).  We have 3~spherical cells with different
internal radii~$R$: 7.0~cm, 3.0~cm, and 1.5~cm. The cylindrical port
tube has internal radius~$r = \q{1.05}{cm}$ and length~$l =
\q{16}{cm}$; its dimensions are the same for all 3~cells. These cells
can be easily interchanged without affecting the diameter of the laser
beams, their alignment, the magnetic field, and the optical detection
system.  The same vacuum conditions can also be obtained, albeit with
different pumping times.

In these experiments we do not use an ionic beam. The neutralizer is
therefore removed and the back port of the cell is connected through a
small valve to a source of Rb atoms, whose temperature can be reduced
to as low as \q{-30}{\degC}\ and controlled by a thermocouple.  The
cell, the port tube and the valves are coated with a PDMS coating by
following the standard procedure described in
Ref.~\cite{Atutov:PRA:1999}. We have also used a Dryfilm coating and
found similar results, but in this paper only the results with PDMS
are presented.

Two coils provide a quadrupole magnetic field with field gradients as
large as \q{20}{G/cm}. The trapping laser is a free-running
Ti:sapphire laser, delivering a maximum power of about 600~mW at a
wavelength of 780~nm.  The laser frequency is scanned across the $5
S_{1/2}$, $F = 3$ to $5 P_{3/2}$, $F' = 4$ Rb transition by changing
the temperature of the aluminum resonator.  This leads to a periodic
appearing and disappearing of the cloud of cold trapped atoms. The
scanning time is 10~s, which is slow enough for the trap to reach
steady state.  A passively-stabilized, free-running diode laser with a
power of 10~mW is tuned to the $5 S_{1/2}$, $F = 2$ to $5 P_{3/2}$,
$F' = 3$ transition for repumping.  The trapping and repumping beams
are superimposed by a polarizing cube and split into 6~beams. Their
diameter is expanded by 6~telescopes and controlled with diaphragms to
select the central and uniform part of the beam.  The maximum beam
diameter is 3~cm.

The fluorescence of both the vapor atoms and the trapped atoms is
collected by lenses and then imaged by a fast and calibrated
photodetector equipped with an iris diaphragm and an interferential
filter, both placed near the surface of the photodetector.  The signal
from atoms in the vapor is usually weaker than the one from trapped
atoms. For this reason, the diameter of the iris diaphragm is chosen
to be large enough to make the two signals comparable.  As a
consequence, the image of the cloud of cold atoms is much smaller than
the size of the photodetector, and its signal is not sensitive to
variations in the position of the trap or in the size of the iris
diaphragm.  The signal is recorded by a lock-in amplifier and a
digital oscilloscope connected to a personal computer.  It is verified
that the electronic response of the detector varies linearly with the
intensity of incident light.

\section{Density of atoms in the vapor cells}
\label{sec:density}

\begin{figure}
\begin{center}
\resizebox{\columnwidth}{!}{\includegraphics{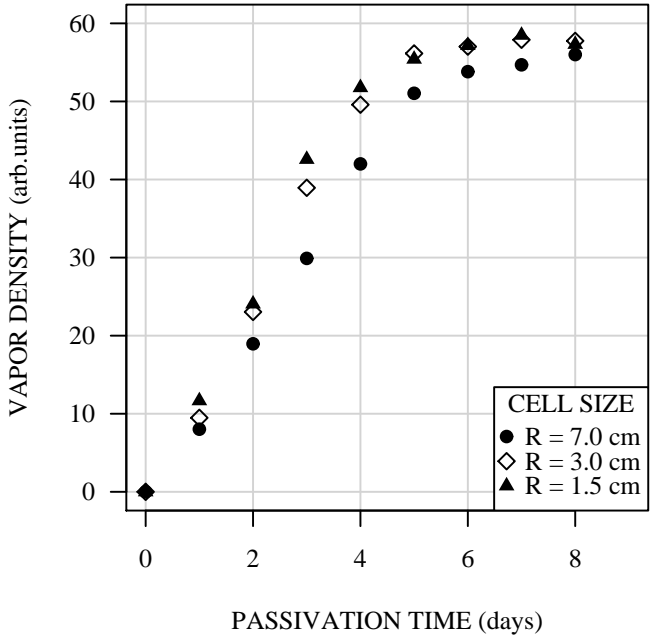}}
\end{center}
\caption{Vapor density versus time of passivation for three different
  cell sizes.}
\label{fig:density_vs_passivation}
\end{figure}

\begin{figure}
\begin{center}
\resizebox{\columnwidth}{!}{\includegraphics{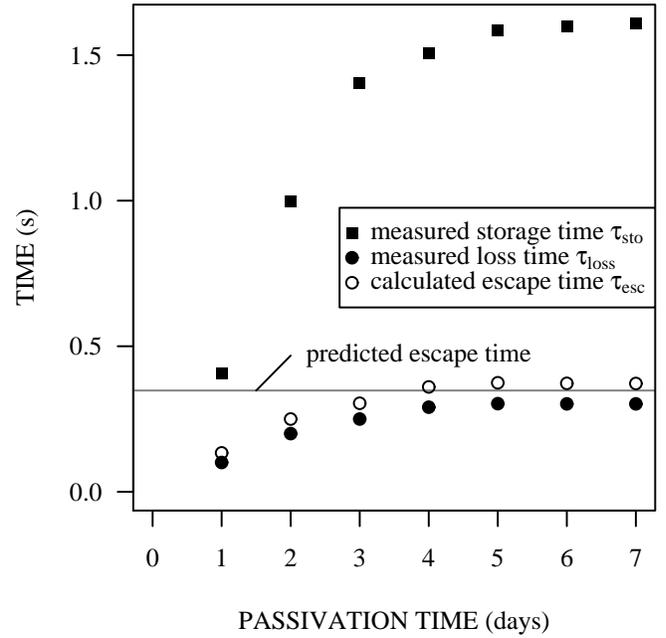}}
\end{center}
\caption{The measured storage time~\tausto, the measured loss
  time~\tauloss, and the resulting escape time~\tauesc\ in the large
  cell versus time of passivation.}
\label{fig:taus_vs_passivation}
\end{figure}

In this section, we present the results of a study on the vapor
density in the 3 different cells when the MOT trapping beams and
magnetic field are turned off. A high density or a large total number
of trappable atoms in the vapor is an important starting point to
achieve high-efficiency magneto-optical trapping of radioactive atoms.

At the beginning of each experiment, we measure the level of stray
light generated by each cell. We find that the intensity of stray
light strongly depends on cell size, and it is roughly
proportional to the inverse cube of the cell radius. The small cell
produces an intensity that is about 50~times larger than that of the
large cell. A high level of stray light can seriously disturb the
detection of small fluorescence signals, both from trapped and vapor
atoms.

We find that all freshly-coated cells do not show any fluorescence
from Rb atoms when valve~VR is opened, meaning that the storage
time~\tausto\ is very small. This can be attributed to the fact that a
fresh coating has a chemically active surface, probably due to
chemically-active gases such as oxygen or water adsorbed on its
surface. Rubidium atoms can also diffuse inside the coating, where
they can be trapped by chemically-active molecules mixed with the
molecules of the coating.

To minimize the residual chemical activity of the coating, we carry
out a passivation (or curing)
procedure~\cite{Lu:PRL:1997,Bouchiat:PR:1966}. First of all, we
continuously pump the cell to obtain a residual-gas pressure of
\q{10^{-8}}{mbar}. It usually takes about one week to get good vacuum
conditions for the largest cell and less time for smaller cells.  To
start passivation, we heat the source of sodium atoms and open
valve~VS, so that the pressure of the alkali vapor in the cell is kept
at about \q{10^{-7}}{mbar}.  We investigate passivation with sodium,
potassium and rubidium itself, obtaining very similar results. In the
following, only measurements with sodium passivation are discussed.

To measure the steady-state density of trappable atoms \nzero, we tune
the laser frequencies to be resonant with the trapping and repumping
transitions of rubidium, until we obtain maximum fluorescence in a
separate reference cell. Five of the laser beams are blocked,
leaving only one open. This is done to avoid the influence of optical
molasses on the escape time and on the atomic density at
equilibrium. Valve~VS is then closed. This prevents sodium from
entering the cell and rubidium from leaking out.  Subsequently,
valve~VR is opened and atoms from the Rb source, kept at constant
temperature, are allowed to fill the cell, until equilibrium is
reached. This usually takes about 30~s. Finally, we record the level
of fluorescence, which is proportional to the steady-state density of
trappable atoms~\nzero.

Figure~\ref{fig:density_vs_passivation} shows how the equilibrium
density \nzero\ depends on the duration of the passivation process for
each of the 3~cells.  At the beginning of the passivation process, the
equilibrium density is very small in all 3~cells. After about 7~days
of continuous passivation, the equilibrium density approaches a limit,
which is practically the same for all 3~cells. This is in agreement
with the model (Eq.~\ref{eq:neq}).  The smallest cell is passivated in
a slightly shorter time. From our measurements we deduce that, for all
cells, the increase in the equilibrium density after passivation is
approximately a factor $10^4$.

The storage time~\tausto\ and the total loss time~\tauloss\ are
measured in the large cell in parallel with the measurement of the
equilibrium density during passivation.  For the measurement of the
storage time~\tausto, we close valve~VS between the cell and ionic
pump, open valve~VR between the Rb source and the cell and fill the
cell with vapor. Then, we rapidly close the source valve~VR and record
the decay of the vapor fluorescence in the cell. We extract the
storage time from the exponential part of the decay curve, when
valve~VR is closed.  To make sure that Rb atoms are permanently
removed from the vapor, we sometimes heat the cell walls and check
that the vapor does not reappear. This indicates that atoms are bonded
to the surface of the coating by chemisorption rather than
physisorbed.

To obtain the loss time~\tauloss\ we fill the cell with vapor, then
we close valve~VR and immediately open valve~VS, allowing the number
of atoms in the vapor to decay by both leaking out of the cell and by
adsorption on the cell walls.  The transmitted intensity as a function
of time is recorded and the loss time is extracted from the decay
curve.

Figure~\ref{fig:taus_vs_passivation} shows~\tausto\ and~\tauloss\ as a
function of the passivation time in the large cell. Both the storage
and the loss time are initially very small. They increase during
passivation and reach their maximum values after about one week. From
the final value of the storage time $\tausto = \q{1.6}{s}$ and
Eq.~\ref{eq:tausto} with $\vavg = \q{2.7\times 10^4}{cm/s}$ and $R =
\q{7.0}{cm}$, one finds that, according to the model, the average
number of bounces before adsorption on the cell walls is $\chi =
2\times 10^4$.

From the measurements of loss time and storage time, it is possible to
estimate the escape time~\tauesc\ using Eq.~\ref{eq:W} with
infinite~\taudec:
\begin{equation}
\tauesc = \left( \frac{1}{\tauloss} - \frac{1}{\tausto} \right)^{-1}.
\end{equation}
After one week of passivation, we measure $\tauloss = \q{0.30}{s}$ and
$\tausto = \q{1.6}{s}$. From these values, we find $\tauesc =
\q{0.37}{s}$. This measurement is in agreement with the prediction
given by Eq.~\ref{eq:tauesc}. In fact, for a cell radius $R =
\q{7}{cm}$, port-tube dimensions $r = \q{1.05}{cm}$ and $l =
\q{16}{cm}$, and Rb thermal velocity $\vavg = \q{2.7\times
  10^4}{cm/s}$, Eq.~\ref{eq:tauesc} yields $\tauesc = \q{0.35}{s}$.

With this technique, it is not possible to measure loss times and
escape times in the smaller cells. These times are short, and the
valves' opening and closing times cannot be neglected. This is also
the reason why the estimated escape time in the large cell
(Fig.~\ref{fig:taus_vs_passivation}) is not constant, as one may
expect. Nevertheless, useful information on the performance of the
large cell can be extracted, as shown above.

\section{Trap population and trapping efficiency}
\label{sec:trapmeas}

\begin{figure}
\begin{center}
\resizebox{\columnwidth}{!}{\includegraphics{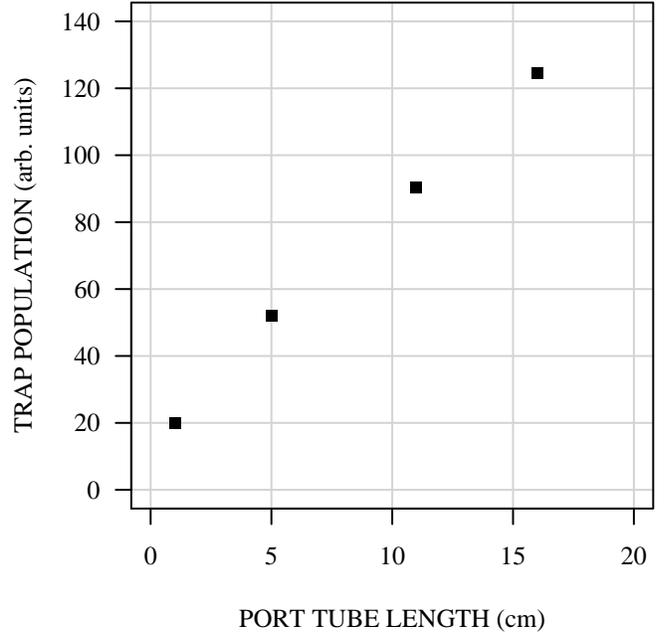}}
\end{center}
\caption{Equilibrium trap population in the large cell versus length
  of the port tube, for a constant incoming flux of Rb atoms.}
\label{fig:trap_vs_tube}
\end{figure}

\begin{figure}
\begin{center}
\resizebox{\columnwidth}{!}{\includegraphics{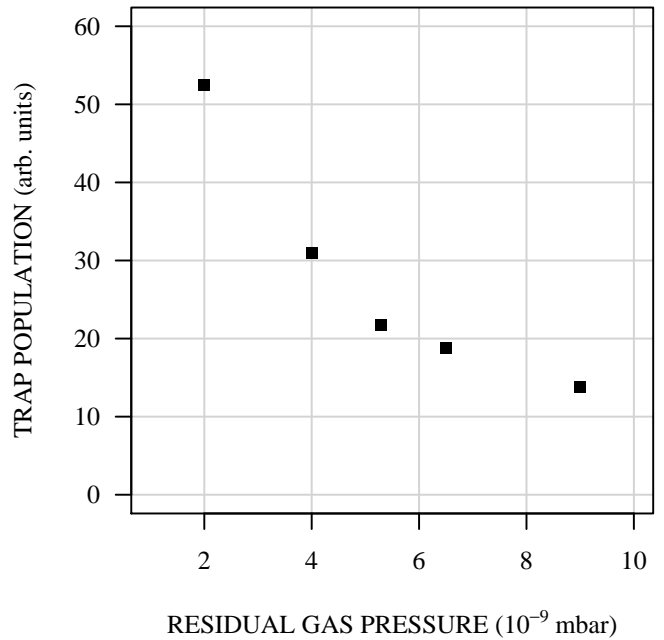}}
\end{center}
\caption{Trap population versus residual gas pressure in the large cell.}
\label{fig:trap_vs_vacuum}
\end{figure}

\begin{figure}
\begin{center}
\resizebox{\columnwidth}{!}{
  \includegraphics{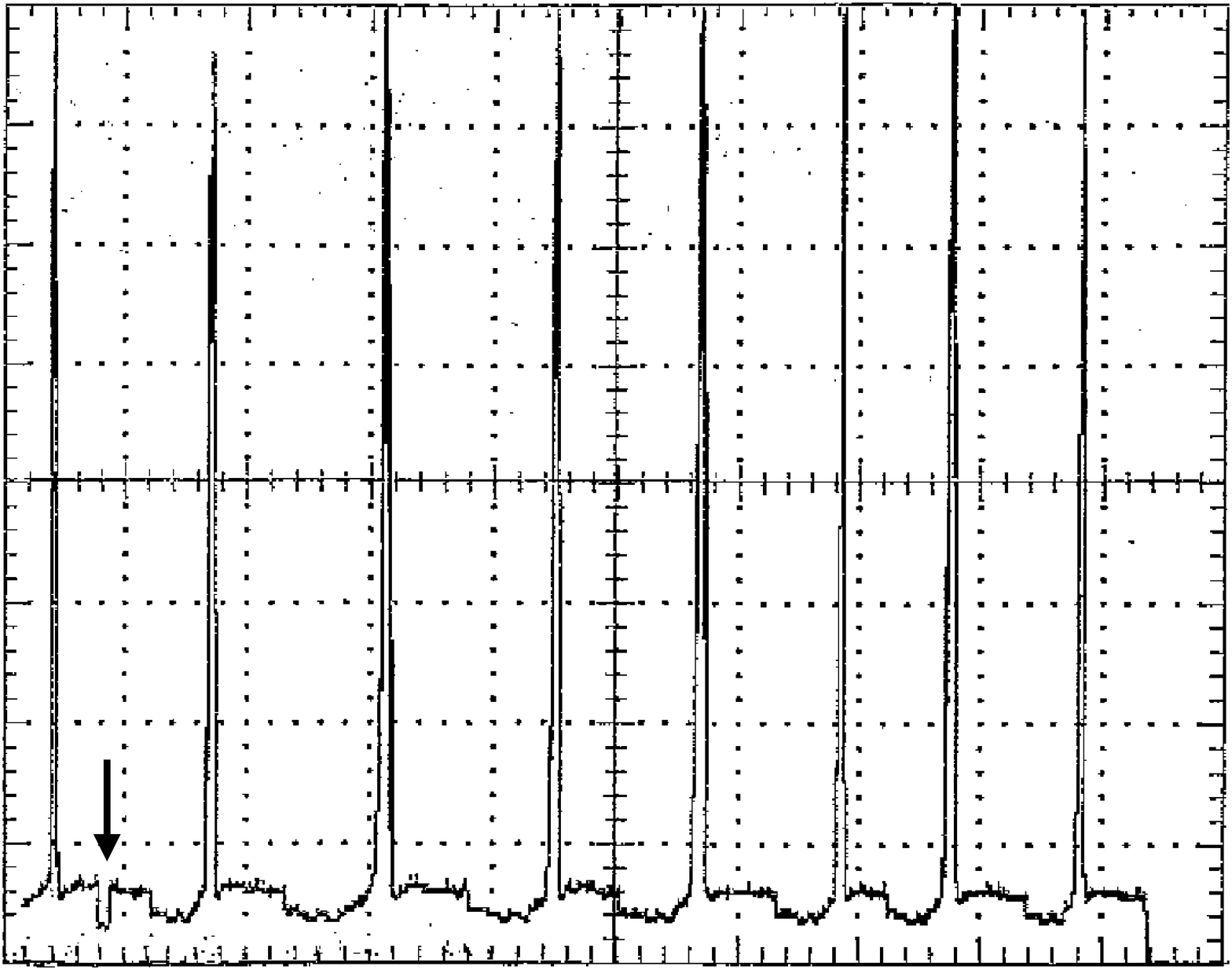}}
\end{center}
\caption{Fluorescence signal as a function of time while scanning the
  frequency of the trapping laser.}
\label{fig:eff}
\end{figure}

\begin{figure}
\begin{center}
\resizebox{\columnwidth}{!}{\includegraphics{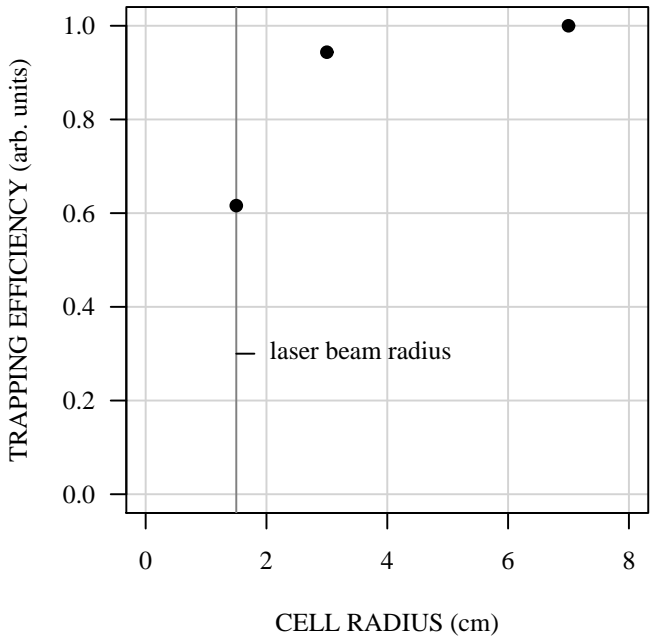}}
\end{center}
\caption{Trapping efficiency versus cell radius.}
\label{fig:eff_vs_size}
\end{figure}

We performed experiments to study the population of cold Rb atoms in
the magneto-optical trap and to measure the trapping efficiency in
3~MOT cells of different sizes.  We detect the fluorescence signal
from cold trapped atoms with the same photodetector and optical system
that is used for the vapor studies.

In all experiments presented in this section, the Rb source is cooled
in order to reduce as much as possible the Rb vapor pressure in the
cell. This is done to minimize the loss of cold atoms due to
collisions with trappable Rb atoms in the vapor. Below a temperature
of about \q{0}{\degC}, we observe that the intensity of fluorescence
from trapped atoms is proportional to the fluorescence from the
vapor. This is an indication that the contribution of collisions
(i.e., a possible term $\propto \Ntrap \Nvap$ in
Eq.~\ref{eq:trapmodel}) is negligible.

In the case of low vapor densities, the model predicts that the trap
population is proportional to the length of the port tube
(Eq.~\ref{eq:Ntraplaser2}). In the large cell, we measure the
variation of the trap population as a function of the length of the
tube by positioning a movable, nitrogen-cooled metallic ring around
the tube itself.  The effective length of the tube is shortened by
adsorption on the cold surface, which acts as a movable atom sink.
With this arrangement, one can smoothly vary the experimental conditions
from a long-port-tube case to a diaphragm case.

Figure~\ref{fig:trap_vs_tube} shows the measured intensity of the trap
fluorescence as a function of the position of the cooled metallic
ring, i.e. the effective length of the port tube.  In accordance with
the model, the trap population appears to increase linearly with
port-tube length.  For the shortest port-tube length of 1~cm, the
equilibrium trap population is about 6~times smaller than that
obtained, in the same cell, with the maximum port-tube length of
16~cm.

To measure the dependence of the trap population on the pressure of
the rest gas in the cell, we switch off the ionic pump attached to the
MOT and let the rest-gas pressure become uniform in the vacuum system.
Without pumping, the rest-gas pressure, which is monitored by a vacuum
gauge positioned near the cell, slowly increases. A measurement of how
the trap fluorescence in the large cell diminishes with increasing
rest-gas pressure is shown in Fig.~\ref{fig:trap_vs_vacuum}.  The plot
shows that the trap population is roughly inversely proportional to
the rest-gas pressure, in agreement with the prediction $\Ntrap
\propto 1/C$ (Eqs.~\ref{eq:Ntrapeq} and~\ref{eq:Ntraplaser2}).

Finally, we provide a comparison of the trapping efficiency measured
in the 3~cells of different sizes. To eliminate uncertainties due to
the Rb flux~$I$, we measure the ratio of two light signals: the
fluorescence intensity from cold atoms in the trap and the one from
trappable atoms in the vapor. Since \nzero\ is proportional to $I$,
the ratio $\Ntrap / \nzero$ is proportional to the trapping
efficiency~$\eta$. We measure this ratio for each cell using the same
optical system and with the same rest-gas pressure.

The fluorescence recorded in the large cell as a function of time is
presented in Fig.~\ref{fig:eff}.  The frequency of the trapping laser
is slowly scanned across the trapping transition several times, while
the frequency of the repumping laser is kept at the maximum of the
repumping line.  In this figure, one can see the fluorescence signal
from trapped atoms (narrow peaks) and from trappable atoms in the
vapor (broad background level). Due to the linear response of the
photodetector, the height of the narrow peaks is proportional to the
number of trapped atoms. To determine a signal proportional to the
density of trappable atoms in the vapor, the frequency of the
repumping laser is briefly detuned away from the repumping
transition. Due to optical pumping, this detuning produces a gap of
zero signal, indicated by an arrow in Fig.~\ref{fig:eff}. The ratio of
signal heights referred to this zero level gives a value that is
proportional to the trapping efficiency~$\eta$.  The value of the
trapping efficiency is obtained by averaging about 10 of these signal
ratios for each cell. The statistical uncertainty is $\pm 1.5\%$. The
trapping efficiency as a function of cell size is shown in
Fig.~\ref{fig:eff_vs_size}. The two larger cells have comparable
efficiency, while the efficiency of the small cell, whose radius
coincides with the radius of the laser beams, is 38\% smaller than
that of the large cell.

The small cell may appear inefficient due to the large curvature of
its walls, which can act as a meniscus lens and affect the spatial
distribution of laser power. This may produce a lower number of
trapped atoms even in the regime of deeply-saturated laser power.  We
check that this possible systematic effect is negligible by inserting
in the laser beams near the cell a thin, milky-white transparent
plastic sheet, which produces a more uniform light distribution.  We
can also use a 6-cm-long, 3-cm-diameter cylindrical cell, which fits
inside the laser beams. In both cases, we obtain the same decrease in
trapping efficiency with respect to the larger cells.

In principle, the resonant ab\-sorp\-tion of light in the trap or in the
vapor changes with the size of the cell, resulting in seemingly
different trapping efficiencies. These systematic effects are
negligible with this technique. In fact, the effect of the optical
thickness of the trap is eliminated by using similar trap populations
in all 3~cells. And even though the optical thickness of the vapor is
larger in larger cells, the fluorescence ratios are not affected,
because trap and vapor fluorescence are attenuated by approximately
the same amount.

These results are in contradiction with the common opinion according
to which the maximum trapping efficiency is associated with a maximum
ratio of laser beam overlap and cell volume.  A possible
interpretation is the following. At first, the optical molasses
collects a sample of cold atoms over the laser beam volume in a few
tens of milliseconds. The cold atoms are then pushed towards the
center of the cell by the relatively weak influence of the magnetic
field in about one second. The magnetic field does not seem to
contribute to the cold atom population. This is suggested by the fact
that the total molasses and trap fluorescence signals are
approximately equal. Atoms in the molasses can drift towards the trap
or be lost due to heating by the cell walls. In the small cell, the
probability of an atom in the molasses to come into contact with the
cell walls and be heated is larger. This interpretation is supported
by the fact that attempts to increase the number of trapped atoms by
frequency chirping, broadband light or by light with several closely
spaced frequency components (to increase the low-velocity tail in the
Boltzmann-Doppler velocity distribution) were not
successful~\cite{Lindquist:PRA:1992,Gibble:OL:1992}.  Experimental
investigations on the role of molasses in a magneto-optical trap will
be presented in a separate paper.

\section{Conclusions}
\label{sec:conclusions}

An experimental study of rubidium magneto-optical traps was presented
and compared with a model of vapor collection and trapping.  We
verified that the density of trappable atoms in the vapor does not
depend on the size of the cell, and demonstrated the importance of
passivating the coated surfaces of the cell. We showed how the trap
population increases with the length of the port tube and with the
quality of the vacuum. A larger trapping efficiency was observed for
the larger cells.

These results are the basis for the design of efficient traps for
radioactive atoms. Using a relatively large cell with a long port tube
has several advantages. Besides being easier to handle, it prolongs
the lifetime of the coating in the presence of an internal heated
neutralizer. A possible disadvantage is the pumping time required to
reach acceptable vacuum conditions, especially in the case of organic
coatings.  For our on-line experiments this is not a problem, because
the pumping time (approximately two weeks) is
anyway shorter than the interval between days of beam time.  Recently,
our MOT was successfully used for the on-line trapping of several
francium isotopes at LNL, INFN's national laboratories in Le\-gna\-ro,
Italy~\cite{deMauro:arXiv:2008,Sanguinetti:arXiv:2008}.

\section{Acknowledgements}

We would like to thank Klaus Jungmann (KVI Groningen) and Carl Wieman
(University of British Columbia and University of Colorado, Boulder)
for reading the man\-u\-script and for sharing their valuable insights.
The authors are also grateful to Paolo Lenisa and Guido Zavattini
(Universit\`a di Ferrara) for their interest in this work and for
useful discussions.

This work was supported by INFN, the Italian institute for nuclear
physics; and MIUR, the Italian governmental department
of education and research.

\bibliographystyle{epj}
\bibliography{Atutov_trap_dyn}   

%
%

\end{document}